\documentclass{aa}  

\usepackage{graphicx}
\usepackage{txfonts}

\usepackage{caption}
\usepackage{subcaption}
\begin{document}

   \title{Effects of non-continuous inverse Compton cooling in blazars}


   \author{A. Dmytriiev
          \inst{1}
          \and
          M. B\"{o}ttcher\inst{1}
          }

   \institute{Centre for Space Research, North-West University, Potchefstroom, 2520, South Africa\\
              \email{anton.dmytriiev@nwu.ac.za}
             }

   \date{Received ...; accepted ...}

 
  \abstract
   {Blazar flares provide a window into the extreme physical processes occurring in relativistic outflows. Most numerical codes used for modeling blazar emission during flares utilize a simplified continuous-loss description of particle cooling due to the inverse Compton (IC) process, neglecting non-continuous (discrete) effects that arise in the Klein-Nishina (KN) regime. The significance of such effects has not been explored in detail yet.}
   {In this study, we investigate the importance of non-continuous Compton cooling losses and their impact on the electron spectrum and spectral energy distribution (SED) of blazars during high flux states (flares), as well as in the low state.}
   {We solve numerically the full transport equation accounting for large relative jumps in energy, by extending our existing blazar flare modeling code EMBLEM. We perform a detailed physical modeling of the brightest $\gamma$-ray flare of the archetypal Flat Spectrum Radio Quasar (FSRQ) 3C\,279 detected in June 2015. We then compare results obtained using the full cooling term and using the continuous-loss approximation.}
   {We show that during flaring states of FSRQs characterized by high Compton dominance, the non-continuous cooling can lead to a significant modification of the electron spectrum, introducing a range of distinct features, such as low-energy tails, hardening/softening, narrow and extended particle excesses, and shifts in the cooling break position. Such distortion translates to differences in the associated SED up to $\sim$ 50\%. This highlights the importance of non-continuous effects and the need to consider them in blazar emission models, particularly applied to extreme $\gamma$-ray flares.}
   {}

   \keywords{radiation mechanisms: non-thermal -- quasars: individual: 3C\,279 -- galaxies: active -- relativistic processes
               }

   \maketitle
%

\section{Introduction}

Blazars, a class of jetted Active Galactic Nuclei (AGN) with the relativistic jet aligned very closely with the line of sight to the Earth, are ideally suited to study the extreme physics of relativistic outflows. These objects emit radiation from the radio band up to the $\gamma$-ray range, with highly variable flux \citep[up to a factor $\sim 10$ or even $\sim 100$, e.g.][]{ackermann2016} across the entire electromagnetic spectrum on time-scales ranging from as short as minutes and hours \citep[e.g.][]{albert2007, hayashida2015} to as long as months and even years \citep[e.g.][]{zacharias2017}. The short time-scale ($t_{\mathrm{var}} \lesssim 1$ week) flux variability, generally referred to as flaring activity, is of particular interest as it enables us to probe the physical processes occurring in blazar jets at their most extreme.

Blazars are categorized as BL Lacertae objects (BL Lac) and Flat Spectrum Radio Quasars (FSRQ), based on the prominence of emission lines in their optical spectra. FSRQs, with noticeable optical emission lines, generally have higher luminosities, primarily dominated by $\gamma$-ray emission. The Spectral Energy Distributions (SED) of blazars typically display two bumps: a low-energy synchrotron radiation bump, and a high-energy bump, the nature of which still remains unclear. In the leptonic scenario this bump is explained as due to the inverse Compton (IC) scattering of low-energy photons by the same electron population producing the synchrotron emission, with the low-energy photons being either synchrotron photons \citep[synchrotron self-Compton (SSC) scenario, e.g.][]{maraschi1992}, or external photons from accretion disk \citep[e.g.][]{dermer1992}, broad line region \citep[BLR, e.g.][]{sikora1994}, and/or dusty torus \citep[e.g.][]{blazejowski2000} (external Compton (EC) scenario). The SSC scenario better explains the SED of BL Lac objects, while the EC scenario is generally preferred for FSRQs \citep[e.g.][]{taveccio1998, abdo2010, boettcher2013}. Additionally, hadronic scenarios have been proposed, in which the $\gamma$-ray emission of blazars is generated via processes involving protons/hadrons \citep[e.g.][]{mucke2001, mucke2003}.

The origin of the blazar flaring behavior remains a mystery despite numerous studies, with several mechanisms proposed to explain this phenomenon. One of the simplest descriptions is a leptonic one-zone model, in which the broad-band emission of blazars originates from a compact region in the jet (a ``blob'') filled with electron-positron plasma and moving relativistically along the jet axis \citep[e.g.][]{katarzynski2001}. Flares arise from perturbations of this configuration, either due to microphysics inside the emitting zone or the macrophysical properties of the blob, such as its geometry or kinematics. In the first case, a flare can be produced due to various transient processes such as particle injection \citep[e.g.][]{albert2007}, and/or acceleration due to shocks \citep[e.g.][]{marschergear1985, sikora2001, boettcher2019}, turbulence \citep[e.g.][]{tammiduffy2009, asano2014, baring2017} or magnetic reconnection \citep[e.g.][]{giannios2009, petrop2016, shuklamannheim2020}. In the second case, emission can be enhanced due to e.g.\ increase of the particle number density or in the Doppler factor of the blob \citep[e.g.][]{casadio2015, paliya2015, larionov2016, luashvili2023}, with the latter caused either by the increase in the bulk Lorentz factor of the blob \citep[e.g.][]{ghisellinitavecchio2008}, or in its viewing angle when the emitting zone moves along a twisted/bent jet \citep{abdobentjet, raiteri2017} or a helical trajectory \citep{villata1999}.

Understanding blazar flaring behavior involves modeling the broad-band emission during flaring states. Two alternative approaches exist to describe the evolution of the emitting particle population (leptonic and/or hadronic) due to various physical processes and/or change of physical conditions in the source: the kinetic \citep[e.g.][]{kardashev1962, kirketal1998} or Monte-Carlo (MC) approach \citep[e.g.][]{mucke1999, summerlinbaring2012}, or the combination of the two \citep[e.g.][]{xuhuichen2011, dimitrakoudis2012}. Various numerical codes developed for blazar modeling implement these approaches. In particular, numerical codes based on the kinetic approach, track the evolution of the spectrum of the particle population residing in the emitting region (and in some models also in the accelerating region) \citep[e.g.][]{mastichiadiskirk1995, chiaberge1999, katarzynski2003, dmytriiev2021} or in the large-scale blazar jet \citep[e.g.][]{zacharias2022}, and calculate the time-dependent particle spectrum evolving due to various physical processes, as well as the associated varying multi-wavelength (MWL) emission. In this work, we specifically focus on the kinetic description of blazar variability within the leptonic framework.

Particles producing the synchrotron and $\gamma$-ray emission lose energy, referred to as particle cooling. Within the leptonic scenario, electrons in the blazar emitting zone suffer energy losses via two mechanisms: (1) synchrotron cooling, and (2) IC processes, in which electrons transfer part of their energy to a photon, upscattering it to high energies (IC cooling). The second mechanism becomes especially important during extreme $\gamma$-ray flares. When the IC process proceeds in the Thomson regime ($\gamma x \ll 1$, where $\gamma$ is the electron Lorentz factor and $x = \epsilon^{\prime}_\mathrm{s} / (m_e c^2)$ is the dimensionless energy of the seed photon), electrons lose only a small fraction of their energy, $\Delta E / E \ll 1$, and the cooling process can be well approximated as a continuous one in the kinetic description. However, when the IC scattering proceeds in the Klein-Nishina (KN) regime ($\gamma x \sim 1$), electrons lose a substantial fraction of their energy in a single interaction, and the process cannot be treated as continuous.

The full transport equation for the latter case has an integro-differential form \citep[e.g.][]{blumgould1970}, presenting a challenge compared to equations in partial derivatives. To remedy this problem, several authors have derived continuous-loss approximations attempting to include KN effects, providing a reasonable description of the IC cooling \citep[e.g.][]{boettcheretal1997, moderski2005}. As a result, the non-continuous effects of cooling have been previously neglected in blazar emission models. This was commonly accepted since \cite{zdziarski1988} have shown that substantial differences in the electron spectrum shape only appear in the case of mono-energetic injection and target photon field spectra. Consequently, most existing blazar emission modeling codes employ a continuous-loss term in the kinetic equation to describe IC cooling, despite its validity being limited to cases with small relative energy losses per scattering event. This is a noteworthy simplification, given that blazars often exhibit significant Klein-Nishina effects.

In this study, we revisit the full integro-differential equation for particle IC cooling in the KN regime that properly treats the large relative jumps of electrons in energy, and investigate deviations from the continuous-loss approximation in the context of blazars. We concentrate on scenarios where discrete cooling effects are most pronounced. Specifically, we examine the flaring states of FSRQs, characterized by strong Compton dominance ($U_{\mathrm{rad}} \gg U_{\mathrm{B}}$), and consider the target photon field of the Broad Line Region (BLR), whose spectral shape closely resembles a monoenergetic distribution. We identify distinctive features induced by the discrete jumps of electrons in energy in the electron distribution and SED, while also assessing the overall significance of such effects in blazars.

\section{Transport equation and numerical approach}

In this section we examine the non-continuous cooling terms in the kinetic equation, and present the numerical implementation for solving it.

\subsection{Transport equation with the full cooling term}

In our model, we assume a one-zone leptonic emission scenario, with electron-positron plasma in the compact region of the jet (a ``blob'') producing the observed varying broad-band emission. We will refer to electrons and positrons as electrons further on. The $\gamma$-ray emission in this scenario is produced via the IC process and the electrons therefore experience IC cooling. A standard form of the kinetic (Fokker-Planck) equation describing the evolution of the electron spectrum $N_e(\gamma,t)$ in the blazar emitting zone due to particle injection, escape and cooling, with the latter treated as a continuous process, is given by \cite[e.g.][]{chiaberge1999}: 

\begin{equation}
 \label{eq:kineticeqstandard}
     \dfrac{\partial N_e(\gamma,t)}{\partial t} = \dfrac{\partial}{\partial \gamma} \left[ - \dot{\gamma}_{\mathrm{cool}} N_e(\gamma,t) \right] - \frac{N_e(\gamma,t)}{t_{\mathrm{esc}}} + Q_{\mathrm{inj}}(\gamma,t),
 \end{equation}

where $t_{\mathrm{esc}}$ is the characteristic time-scale of particle escape, $Q_{\mathrm{inj}}(\gamma,t)$ is the spectrum of injected particles, and the quantity $\dot{\gamma}_{\mathrm{cool}}$ is the continuous cooling rate, which is a sum of synchrotron and IC cooling rates, $\dot{\gamma}_{\mathrm{cool,syn}}$ and $\dot{\gamma}_{\mathrm{cool,IC}}$ respectively:

\begin{gather}
\label{eq:coolingrates}
    \dot{\gamma}_{\mathrm{cool}} = \dot{\gamma}_{\mathrm{cool,syn}} + \dot{\gamma}_{\mathrm{cool,IC}} \\
    \dot{\gamma}_{\mathrm{cool,syn}} = - \frac{4 \sigma_{\mathrm{T}} U_B}{3 m_e c} \gamma^2
\end{gather}
 
with $U_B = B^2/(8 \pi)$ is the magnetic field density, $\sigma_{\mathrm{T}}$ is the Thomson cross-section, $m_e$ is the electron rest mass, and $c$ is the velocity of light in vacuum. For the IC cooling rate $\dot{\gamma}_{\mathrm{cool,IC}}$, a useful continuous-loss approximation was derived by \cite{moderski2005}:

 \begin{equation}
 \label{eq:continuouscoolrate}
 \dot{\gamma}_{\mathrm{cool,IC}} \, = \, - \frac{4 \sigma_\mathrm{T}}{3 m_e c} \, \gamma^2 \, \int_{x_{\mathrm{min}}}^{x_{\mathrm{max}}} f_{\mathrm{KN}}(4\gamma x) u^{\prime}_{\mathrm{rad}}(x) dx \,
 \end{equation}
 
with
 
 \begin{equation}
f_{\mathrm{KN}}(z) =
    \begin{cases}
        (1+z)^{-1.5}, & \text{for } z < 10^4\\
        \frac{9}{2z^2} \left(\text{ln}(z) - \frac{11}{6} \right), & \text{for } z \geq 10^4
    \end{cases}
\end{equation}

where $u^{\prime}_{\mathrm{rad}}(x)$ represents the energy density of the target photon field (per unit of photon energy interval), with both energy density and target photon energy $x$ being in the frame of the blob, and $x_{\mathrm{min/max}}$ being minimum and maximum dimensionless energies of soft photons, respectively. Despite an attempt at reasonably treating the KN effects, in particular, the cross-section decrease with increasing energy of the seed photon in the KN regime, the above-mentioned approximation cannot accurately take into account large jumps of particles in energy due to the inherent non-continuous nature of this effect. The full transport equation to treat large jumps of particles in energy in an exact manner may be written as \citep[e.g.][]{blumgould1970, zdziarski1988}:
 
 \begin{multline}
 \label{eq:fulltransport}
     \dfrac{\partial N_e(\gamma,t)}{\partial t} = - N_e(\gamma,t) \int_{1}^{\gamma} C(\gamma,\gamma^{\prime}) d\gamma^{\prime} + \int_{\gamma}^{\infty} N(\gamma^{\prime},t) C(\gamma^{\prime},\gamma) d\gamma^{\prime} + \\ 
     + \dfrac{\partial}{\partial \gamma} \left[ - \dot{\gamma}_{\mathrm{cool,syn}} N_e(\gamma,t) \right] - \frac{N_e(\gamma,t)}{t_{\mathrm{esc}}} + Q_{\mathrm{inj}}(\gamma,t)
 \end{multline}
 
with

 \begin{equation}
\label{eq:Cdefinition}
 C(\gamma,\gamma^{\prime}) = \int_{E_{*}/\gamma}^{\infty} dx \, n^{\prime}_0(x) \, \frac{3 \sigma_\mathrm{T} c}{4 E \gamma} \left[ r + (2-r) \, \chi - 2\chi^2 + 2\chi \text{ln} \chi \right]
 \end{equation}
 
 being the Compton kernel by \cite{jones1968}, expressing the rate (per unit of time) of electron downscattering from $\gamma$ to $\gamma^{\prime}$, $n^{\prime}_0(x)$ being the number density of the target photons (per unit of photon energy interval) in the frame of the blob, and
 
\begin{gather}
x = \frac{\epsilon_\mathrm{s}}{m_e c^2} \, , \hspace{3mm} \chi = \frac{E_{*}}{E} \, , \hspace{3mm} E = \gamma x \\
E_{*} = \frac{1}{4} (\gamma/\gamma^{\prime} - 1) \, , \hspace{3mm} E > E_{*} \\ 
r = \frac{1}{2} (\gamma/\gamma^{\prime} + \gamma^{\prime}/\gamma)
\end{gather}

The first integral term on the RHS of Eq.~\ref{eq:fulltransport} represents IC downscattering from $\gamma$ to lower Lorentz factors, while the second integral term is describing IC downscattering from higher Lorentz factors to $\gamma$. The synchrotron losses are still represented by a continuous-loss term $\dot{\gamma}_{\mathrm{cool,syn}}$, which is reasonable given the (always) small fractional energy loss in this process.

\subsection{Numerical implementation}

To solve the full transport equation (Eq.~\ref{eq:fulltransport}), we use and extend the existing numerical code \texttt{EMBLEM} by \cite{dmytriiev2021}, which is solving the standard kinetic equation Eq.~\ref{eq:kineticeqstandard} (with additional Fermi-I and Fermi-II acceleration terms) using the \cite{changcooper1970} numerical scheme, suited for the above-mentioned equation form, in particular, with energy losses given by a continuous term. A number of modifications were introduced to the \texttt{EMBLEM} code to adapt it for solving the equation with a full cooling term.

The Compton kernel (Eq.~\ref{eq:Cdefinition}) is sharply peaked around the region $\gamma \approx \gamma^{\prime}$, which makes numerical integration challenging. We follow the approach proposed by \cite{zdziarski1988} where the integration domain spanning from $\gamma = 1$ up to $\gamma \rightarrow \infty$ is split into three sub-regions, with the middle peculiar region $\gamma/(1+\delta) \leq \gamma^{\prime} \leq \gamma (1+\delta)$, $\delta \ll 1$, being treated separately. By performing a Taylor expansion around the peculiar point $\gamma = \gamma^{\prime}$, the term for the middle region can be reduced to a continuous-loss form (same form as the first term in the RHS of Eq.~\ref{eq:kineticeqstandard}). Following this approach, the full IC cooling term is then expressed via three different terms \citep{zdziarski1988}:

\begin{multline}
    \label{eq:threeregions}
    \left[\dfrac{\partial N_e(\gamma,t)}{\partial t}\right]_{\mathrm{cool, IC}} = - N_e(\gamma,t) \int_{1}^{\gamma/(1+\delta)} C(\gamma,\gamma^{\prime}) d\gamma^{\prime} \, + \\ +  \, \dfrac{\partial}{\partial \gamma} \left[- \dot{\gamma}_{\mathrm{CL,IC}} N_e(\gamma,t) \, \right] \, + \, \int_{\gamma (1+\delta)}^{\infty} N_e(\gamma^{\prime},t) C(\gamma^{\prime},\gamma) d\gamma^{\prime}
\end{multline}

with $\dot{\gamma}_{\mathrm{CL,IC}}$ representing the IC cooling transition rate in continuous-loss regime:

\begin{equation}
    \label{eq:continuouslosspart}
    \dot{\gamma}_{\mathrm{CL,IC}} = - \int_{\gamma/(1+\delta)}^{\gamma} C(\gamma,\gamma^{\prime}) (\gamma - \gamma^{\prime}) d\gamma^{\prime}
\end{equation}

It is worth noting that the rate $\dot{\gamma}_{\mathrm{CL,IC}}$ is not the same as $\dot{\gamma}_{\mathrm{cool,IC}}$ (Eq.~\ref{eq:continuouscoolrate}), as the latter attempts to approximate the combined effect provided by all three terms of Eq.~\ref{eq:threeregions} in a continuous framework, while the rate $\dot{\gamma}_{\mathrm{CL,IC}}$ only incorporates small energy losses suffered by electrons. To avoid troublesome numerical integration with a sharply peaked integrand, the Eq.~\ref{eq:continuouslosspart} is integrated analytically over the Lorentz factors \citep{zdziarski1988}, which yields:

\begin{multline}
    \label{eq:contcoolanalytical}
    \dot{\gamma}_{\mathrm{CL,IC}} \approx - \int_{x_{\mathrm{min}}}^{x_{\mathrm{max}}} dx \, n^{\prime}_0(x) \sigma_{\mathrm{T}} \gamma c s g^2 \left[\dfrac{3}{2} + \dfrac{g}{3} + 2 g \, \text{ln}(g) \, - \right. \\ \left. - \, \dfrac{3}{2} g^2 - 9 s g \left(\dfrac{1}{3} + \dfrac{g}{8} + \dfrac{g}{2} \text{ln}(g) - \dfrac{2}{5} g^2 \right)  \right],
\end{multline}

where $s = 4 x \gamma$, and $g = \text{min}(\delta/s, 1)$.

To adapt the \texttt{EMBLEM} code to the integro-differential transport equation (Eq.~\ref{eq:fulltransport}), we first transform it into a form where the terms have the same mathematical structure as the standard kinetic equation (Eq.~\ref{eq:kineticeqstandard}) that the original code is designed to solve, relying on the \cite{changcooper1970} numerical scheme. This is achieved by numerically computing at each time step the non-continuous cooling terms in Eq.~\ref{eq:threeregions} on the Lorentz factor grid, substituting the sought quantity $N_e(\gamma,t)$ with the electron spectrum from the previous time step (ensuring a sufficiently small time step on the time grid). The result represents a certain tabulated function of $\gamma$ and can therefore be treated as a time-dependent source (injection-like) term additional to $Q_{\mathrm{inj}}(\gamma)$. For the continuous part of the full cooling term (Eq.~\ref{eq:threeregions}), we compute the IC cooling rate in the continuous regime $\dot{\gamma}_{\mathrm{CL,IC}}$ using Eq.~\ref{eq:contcoolanalytical}. The integral over the seed photon energies is evaluated numerically and the result is added to the synchrotron cooling rate (Eq.~\ref{eq:coolingrates}) in the code, yielding the total continuous cooling rate, treated in the same manner as $\dot{\gamma}_{\mathrm{cool}}$ in Eq.~\ref{eq:kineticeqstandard}. As a result, we simplify the integro-differential equation to a standard differential one, enabling the application of the \cite{changcooper1970} scheme. After solving the equation, we obtain the updated electron spectrum, which is then used to reevaluate the non-continuous cooling terms on the Lorentz factor grid, improving the accuracy of the estimated values. This process is repeated iteratively until convergence is achieved. Throughout our numerical computations, we use $\delta = 0.05$. For each solution, we ensure that the normalizations (total number of particles) are equal between the electron spectra for non-continuous and continuous-loss cooling scenarios.

\section{Application to the FSRQ 3C\,279}

In this section we apply the resulting code to simulate the electron spectrum and SED during a realistic flaring state, representing a recent flare of the FSRQ 3C\,279, and make a comparison of the results with continuous versus non-continuous cooling case.

\subsection{3C\,279: the archetypal FSRQ}

3C\,279 ($z = 0.5362$) is a prototypical FSRQ that exhibits intense and complex variability across all wavebands, including the $\gamma$-ray range. It is one of the most studied blazars thanks to numerous MWL campaigns on the target \citep{hayashida2012, hayashida2015}. The multi-frequency SED of 3C\,279 displays a characteristic two-bump structure with the low-energy component peaking in the infrared band, and the high-energy component reaching its maximum in the domain $\sim$0.1 GeV -- $\sim$ a few GeV, depending on the spectral state. 3C\,279 has been continuously observed by the {\it Fermi} Gamma-Ray Space Telescope since 2008. During a high state in 2013 -- 2014, the source showed a very hard $\gamma$-ray spectrum with a high level of Compton dominance $\sim 300$, and a peak $\gamma$-ray flux of $f_{\gamma}(> 0.1 \text{ GeV}) \simeq 10^{-5}$ ph cm$^{-2}$ s$^{-1}$. In June 2015, it showed an extreme flare with the historically highest peak $\gamma$-ray flux, $f_{\gamma}(> 0.1 \text{ GeV}) \simeq (3.6 \pm 0.2) \times 10^{-5}$ ph cm$^{-2}$ s$^{-1}$, and with a very short flux-doubling time-scale of $\lesssim 5$ min \citep{ackermann2016}.

\subsection{Physical scenario for the low and flaring state}

\subsubsection{General model}

For our study, we choose the most extreme flare of 3C\,279, specifically the June 2015 one with the historically highest $\gamma$-ray flux $F_{\gamma}(\text{0.1 -- 100 GeV}) \sim 2 \times 10^{-8}$ erg cm$^{-2}$ s$^{-1}$, as well as the highest ever Compton dominance level of $\sim 800$ \citep[Fig.1 in][]{dmytriiev2023}. We adhere to the one-zone leptonic scenario, assuming that the varying emission of the source is produced within a spherical ``blob'' of a fixed radius $R_{\mathrm{b}}$ that is situated at a distance $r_{\mathrm{emz}}$ from the central black hole and is traveling down the jet at a velocity close to the speed of light, characterized by a bulk Lorentz factor $\Gamma$ and a Doppler factor $\delta_{\mathrm{b}}$. To reduce the number of free parameters, we assume the viewing angle of the jet $i_{\rm j}$ such that $\delta_{\rm b} = \Gamma$ at all times. The $\gamma$-rays are produced in the blob via IC scattering of a target photon field. This target photon field comprises two components: an external field of the BLR (time-independent) and the synchrotron radiation field (time-dependent). In our case of FSRQs, the external photon field dominates the high-energy emission production and IC cooling processes. We model the BLR photon field simply as a single Gaussian-shaped Ly $\alpha$ emission line with a fixed central energy $E_{\mathrm{Ly} \alpha \mathrm{, c}} = 10.2$ eV and fixed width $\Delta E_{\mathrm{Ly} \alpha} = 0.2$ eV corresponding to an average velocity of BLR clouds of $\sim 6000$ km s$^{-1}$ (both energies given in the BLR frame). The luminosity of the BLR is considered to be a constant fraction $\xi_{\mathrm{BLR}}$ of the accretion disk luminosity, for which we adopt the value $L_{\mathrm{AD}} = 6 \times 10^{45}$ erg s$^{-1}$ \citep{hayashida2015}. The calculation of the energy density of the external photon field in the frame of the blob follows the approach by \cite{hayashida2012}. We assume that particles are injected into the blob with a log-parabola electron injection spectrum above a certain minimum injection Lorentz factor $\gamma_{\mathrm{inj,c}}$,

 \begin{equation}
 Q_{\mathrm{inj}}(\gamma) =
    \begin{cases}
        K_{\mathrm{inj}} \left(\dfrac{\gamma}{\gamma_0} \right)^{-\alpha_{\mathrm{inj}} - \beta_{\mathrm{curv}} \text{log10}(\gamma/\gamma_0)}, & \text{for } \gamma \geq \gamma_{\mathrm{inj,c}}\\
        0 , & \text{for } \gamma < \gamma_{\mathrm{inj,c}}
    \end{cases}
\end{equation}

with $\gamma_0 = 200$ being the (arbitrarily chosen) pivot Lorentz factor. We attribute the lower-energy cutoff $\gamma_{\mathrm{inj,c}}$ in the injection spectrum to the electron-proton co-acceleration occurring in the shocked region of the jet \citep[refer to e.g.,][]{zechlemoine2021}; however, we do not model this process in the present study. It is worth mentioning that we also attempted the modeling using a power law injection spectrum with and without an exponential cutoff. However, we were not able to reproduce the low-state SED data with such injection spectra. Finally, particles also escape the emitting region at a characteristic time-scale of $t_{\mathrm{esc}} \sim 1 \ R_{\mathrm{b}}/c$. Throughout this work, we use a redshift of 3C\,279 of $z = 0.5362$, and the Hubble parameter value of $H_0 = 70$ km s$^{-1}$ Mpc$^{-1}$.

\subsubsection{Low state model}

Following this framework, we first model the low state of the source using the multi-band data set from \cite{hayashida2012}. We simulate the low state as an asymptotic steady-state established due to the competition of injection, cooling and particle escape \cite[based on the approach by e.g.][]{boettcher2013, dmytriiev2021}. In this case, softening of the electron spectrum due to cooling arises naturally without a need to artificially introduce this effect. In addition, in our modeling, we disregard the low-energy radio measurements, assuming that emission from the source in this regime is dominated by the extended jet, rather than the compact emitting zone.

\setlength{\tabcolsep}{5.4pt}
\begin{table*}[t]
\caption{Low-state modeling parameters. \label{tab:lowstatefit}
}
\vskip 0cm
\centering

\begin{tabular}{cccccccccccc}
\hline
$B$ (G) & $R_\mathrm{b}$ (cm) & $\delta_{\mathrm{b}}$ & $K_{\mathrm{inj}}$ (cm$^{-3}$ s$^{-1}$) &  $\alpha_{\mathrm{inj}}$ & $\beta_{\mathrm{curv}}$ & $\gamma_{\mathrm{inj,c}}$ & $\xi_{\mathrm{BLR}}$ & $r_{\mathrm{emz}}$ (cm) & $E_{\mathrm{Ly} \alpha \mathrm{, c}}$ (eV) & $\Delta E_{\mathrm{Ly} \alpha}$ (eV)  \\
\hline
$1.7$ & $2.7 \times 10^{15}$ & $30$ & $2.8 \times 10^{-4}$ & $2.6$ & $0.35$ & $130$ & $0.1$ & $6.8 \times 10^{17}$ & $10.2$ (f) & $0.2$ (f) \\
\hline
\end{tabular}\\

\vspace{2mm}

{\it Notes:} {The notation ``(f)'' indicates that the parameter value was fixed in the model, and was not varied. The bulk Lorentz factor $\Gamma$ is set to be equal to $\delta_{\rm b}$, which implies a jet viewing angle $i_{\rm j} = 1.9^{\circ}$.} 
\end{table*}

\begin{figure*}[t]
     \centering
     \begin{subfigure}[b]{0.49\textwidth}
         \centering
         \includegraphics[width=\textwidth]{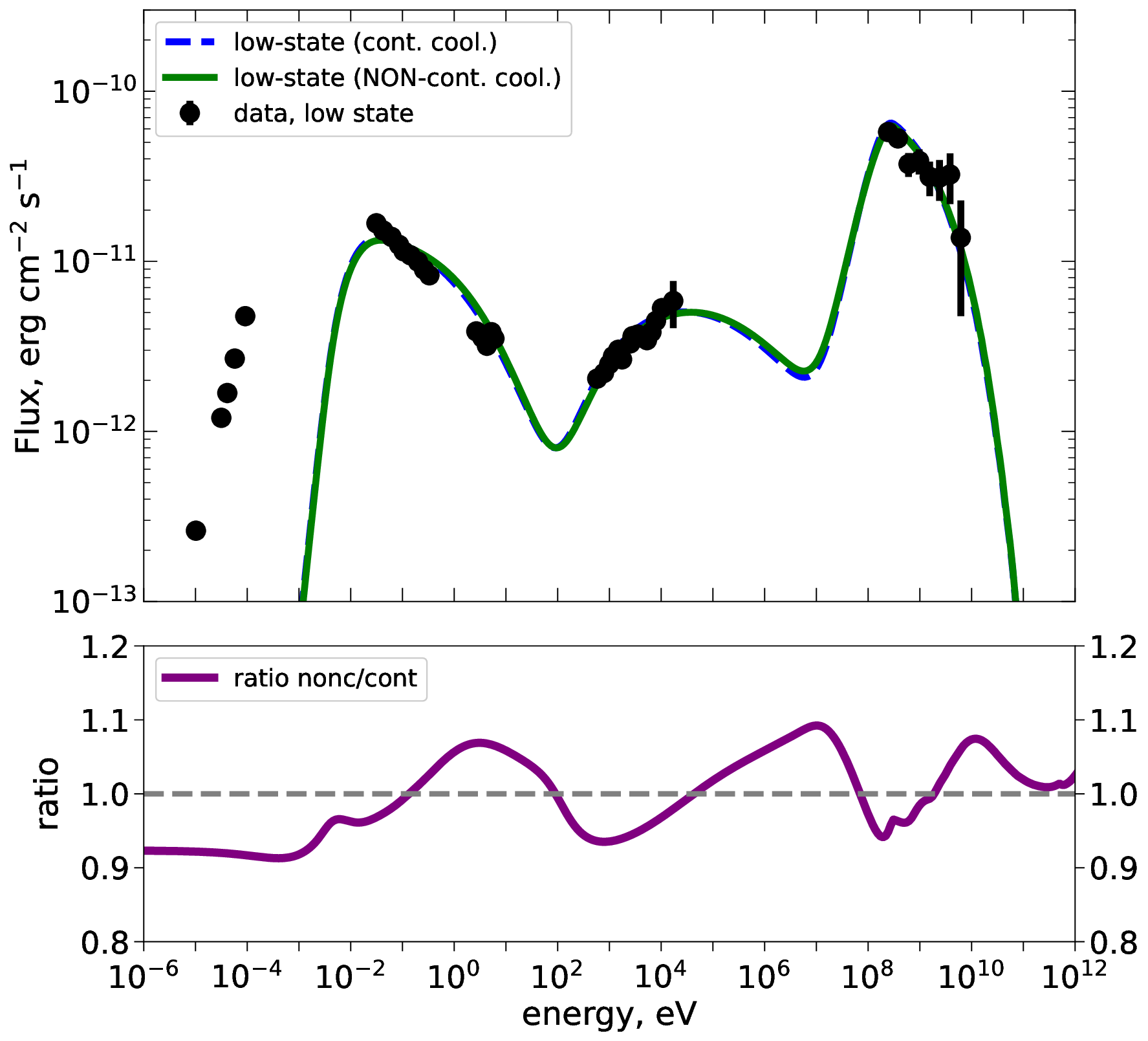}
         \caption{SED}
         \label{fig:lowsed}
     \end{subfigure}
     \hfill
     \begin{subfigure}[b]{0.49\textwidth}
         \centering
         \includegraphics[width=\textwidth]{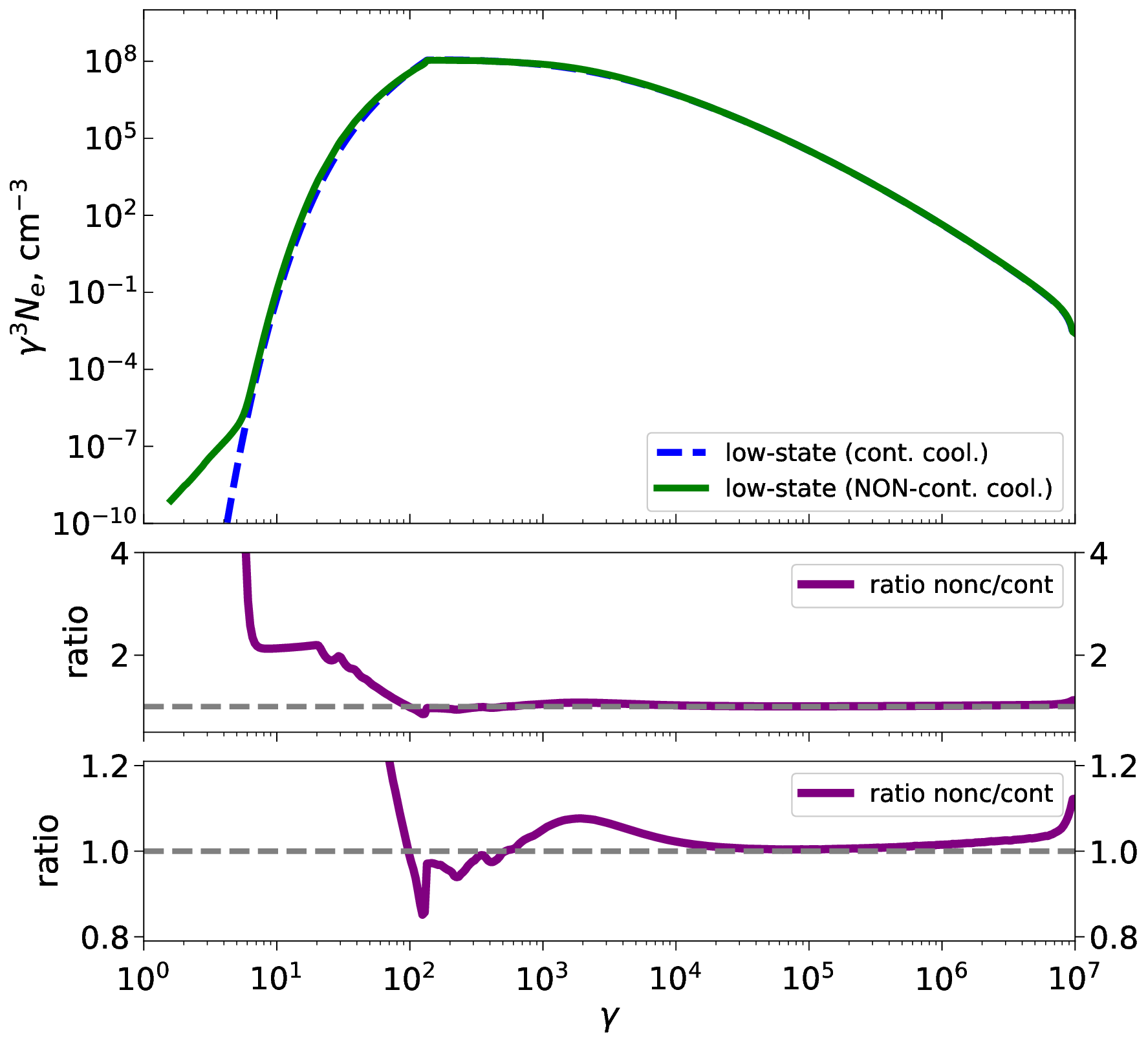}
         \caption{Electron spectrum}
         \label{fig:lowne}
     \end{subfigure}
        \caption{Comparison of the low-state SEDs and electron spectra for the two cooling descriptions. The blue dashed curves represent the model with continuous-loss approximation for cooling \citep{moderski2005}, whereas green solid curves indicate the same model, but with the full non-continuous cooling term. Left panel: modeling of the low state SED of 3C\,279. The black points display the low-state data set of the source \citep{hayashida2012}. The adjacent narrow bottom panel provides a zoom into the ratio of the SED for non-continuous to continuous-loss cooling description (in linear scale). Right panel: the underlying low-state electron spectra for the two cooling descriptions. The electron spectra are shown in $\gamma^3 N_e(\gamma)$ representation to better highlight the discrepancies between the two cooling models. Adjacent narrow bottom panels visualize the ratio of the electron spectrum for non-continuous to continuous-loss cooling scenario (in linear scale), with the second lower panel zooming into the Lorentz factor domain above 100.}
        \label{fig:low}
\end{figure*}

\subsubsection{June 2015 flare model}

Next, we perform modeling of the June 2015 flaring state. We use the MWL data collected during the dedicated observational campaign, which includes spectral measurements in the optical-UV (UVOT instrument), X-ray ({\it Swift}-XRT), $\gamma$-ray ({\it Fermi}-LAT) and Very High Energy (VHE) $\gamma$-ray (H.E.S.S.) bands, reported by \cite{hess2019flare3c279}. Our primary focus is on accurately reproducing the SED at the flare peak (the ``Maximum'' time frame in \cite{hess2019flare3c279}, MJD 57189.125 -- 57189.25), while also attempting to provide a satisfactory fit of the SED $\sim 8.9$ h prior to the peak at the early stage of the outburst (the ``Night 1'' time frame in \cite{hess2019flare3c279}, MJD 57188.756 -- 57188.88) within the self-consistent time-dependent framework. However, our modeling approach does not include fitting the shapes of MWL light curves. Despite this, our methodology ensures a reasonable approximation of the dynamic time-scales around the flare peak.

Following the framework introduced by \cite{dmytriiev2021}, we treat the flaring state as a perturbation of the low-state configuration, induced by a transient physical process (e.g.\ particle acceleration, injection, etc.), which may be accompanied by a change in one or a few global physical parameters. To minimize complexity, we adhere to the one-zone scenario with transient Fermi-II re-acceleration of particle population within the blob due to intervening turbulence, simulated using a dedicated setup of the \texttt{EMBLEM} code (see Section~6.1 of \cite{dmytriiev2021}). We vary the Fermi-II acceleration time-scale $t_{\rm acc,FII}$ and the escape time-scale $t_{\rm esc,FII}$, with the latter typically longer than the low-state escape time-scale due to particle diffusion through the turbulent magnetic field \citep[e.g.,][]{tramacere2011}. Another parameter is the total duration of the perturbative acceleration episode $t_{\rm dur,FII}$, which is constrained based on the time span of the $\gamma$-ray flux rise observed in the {\it Fermi}-LAT light curve of the flare \citep{hess2019flare3c279}, $\sim 12$ h.

We also assess the significance of the triplet pair production (TPP) process ($e^{-} + \gamma \rightarrow e^{-} + e^{-} + e^{+}$) under our conditions. At high electron and/or target photon energies, TPP losses start competing with IC cooling losses \citep[e.g.][]{mastichiadis1994}. Using developments by \cite{mastichiadis1994}, we determine that for electron interactions with BLR target photons, boosted in the emitting zone frame with a Doppler factor $\delta_{\rm b} \sim$ 10 -- 100, TPP energy losses ($<\dot{\gamma}>$) dominate over IC losses at $\gamma > 10^9$. Since no electrons with Lorentz factors exceeding $\sim 10^7$ exist in the emitting region, we conclude that the TPP loss process has a negligible effect on the electron spectrum in our case.
 
\subsection{Results}

\setlength{\tabcolsep}{9.0pt}
\begin{table}[t]
\caption{ June 2015 flaring state modeling parameters. \label{tab:flaringstatefit}
}
\vskip 0cm
\centering

\begin{tabular}{ccccc}
\hline
$t_{\rm acc,FII}$ & $t_{\rm esc,FII}$ & $\delta_{\rm b,fl}$ & $B_{\rm fl}$ (G) & $t_{\rm dur,FII}$\\
\hline
$3.6 \ R_{\rm b}/c$ & $4.2 \ R_{\rm b}/c$ & 40 & 0.2 & $12.8 \ R_{\rm b}/c$\\
\hline
\end{tabular}\\

\vspace{2mm}

{\it Notes:} The bulk Lorentz factor $\Gamma_{\rm fl}$ is set to be equal to $\delta_{\rm b,fl}$, which implies a jet viewing angle $i_{\rm j,fl} = 1.4^{\circ}$. In the observer's frame, the duration of re-acceleration $t_{\rm dur,FII} = 12.8 \ R_{\rm b}/c$ corresponds to $\approx 12.3$ h.

\end{table}

\begin{figure*}[t]
     \centering
         \includegraphics[width=0.65\textwidth]{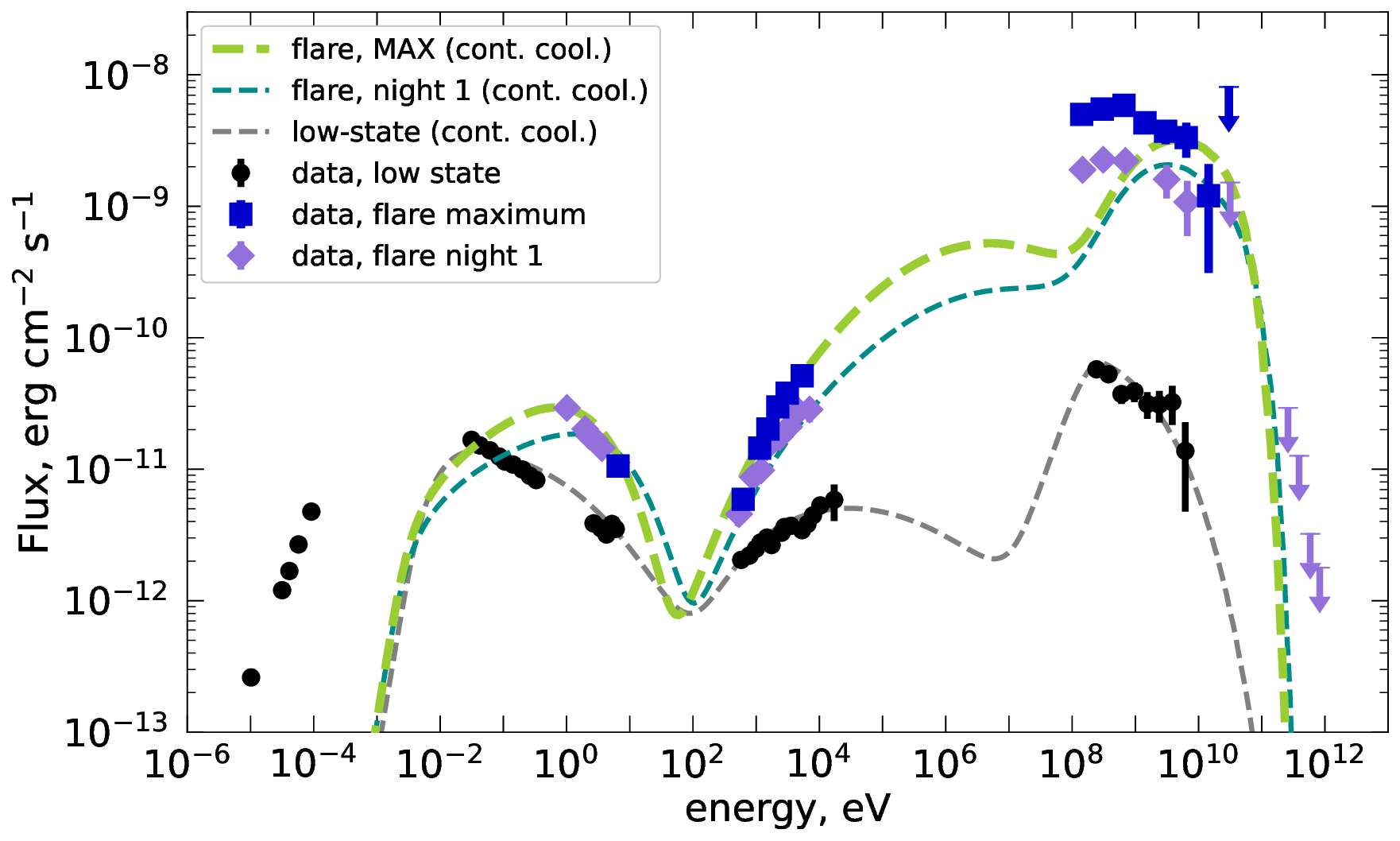}
        \caption{Modeling of the June 2015 flaring state of 3C\,279. The black, blue and purple data points display the MWL data during the low state, flare peak and in the pre-peak state ``Night 1'' ($\approx 8.9$ h earlier), respectively \citep{hess2019flare3c279}. The data points at highest energies are uncorrected for the Extragalactic Background Light (EBL) absorption. The error bars for the optical and X-ray data points are relatively small and are not visible. The gray curve shows the low-state model from Fig.~\ref{fig:lowsed}, the cyan curve indicates the time-dependent flare model at the moment $t_1 = 3.6 \ R_{\rm b}/c$ fitting the pre-peak state data, and the green curve illustrates the same model at the moment $t_2 = 12.8 \ R_{\rm b}/c$ fitting the flare peak data. The model SEDs were absorbed on the EBL using the model by \cite{dominguez2011}. The model curves are displayed in the dashed style to highlight the use of continuous-loss approximation in the model calculation.}
        \label{fig:modelingflare}
\end{figure*}

Initially, we use the continuous-loss approximation for cooling by \cite{moderski2005} to reproduce the observed characteristics of both the quiescent and flaring state. Our model successfully replicates the low-state data set, as illustrated in the Fig.~\ref{fig:lowsed}, with the corresponding model parameters listed in Table~\ref{tab:lowstatefit}. The underlying electron spectrum is depicted in Fig.~\ref{fig:lowne} (blue dashed curves). 

For the flaring state, initially, we attempt to replicate the MWL SED data of the flare peak by adjusting $t_{\rm acc,FII}$ and $t_{\rm esc,FII}$, but find that the model consistently underpredicts the observed $\gamma$-ray flux level and Compton dominance. To address this, we also vary the Doppler factor $\delta_{\rm b, fl}$ (with the bulk Lorentz factor again adhering to $\Gamma_{\rm fl} = \delta_{\rm b, fl}$) and magnetic field $B_{\rm fl}$ during the acceleration episode, given that $\Gamma_{\rm fl}$ and $B_{\rm fl}$ control the Compton dominance, $CD \propto U_{\rm rad} \Gamma_{\rm fl}^2 / B_{\rm fl}^2$, with $U_{\rm rad}$ being the density of the BLR field in the AGN rest frame. The bulk Lorentz factor may be enhanced due to passage of the blob through an active part of the jet \citep[e.g.][]{ghisellinitavecchio2008}, with the Doppler factor increasing to reach $\delta_{\rm b, fl} = \Gamma_{\rm fl}$ owing to a modest decrease in the jet viewing angle due to e.g.\ the curvature of this jet segment. The magnetic field decrease may arise from magnetic reconnection and/or turbulent dissipation of magnetic energy in this jet region. Furthermore, a study by \cite{dmytriiev2023} concludes that the states of extreme Compton dominance in 3C\,279 are most likely explained by strong variations in the magnetic field and moderate variations in the bulk Lorentz factor. The (total) perturbation duration in the blob frame is well constrained and is adjusted only in a narrow range $t_{\rm dur,FII} = (12 \pm 1) \ \text{h} \times \delta_{\rm b, fl} / (1+z)$.

As a result, we find $t_{\rm acc,FII} = 3.6 \ R_{\rm b}/c$, $t_{\rm esc,FII} = 4.2 \ R_{\rm b}/c$, $\delta_{\rm b, fl} = 40$ and $B_{\rm fl} = 0.2$ G, necessary to adequately fit the flare peak data, as well as the pre-peak state (see Table~\ref{tab:flaringstatefit}). The acceleration and escape time-scales appear to be in overall agreement with the doubling and halving time-scales observed in the {\it Fermi}-LAT light curve. The ``Night 1'' (pre-peak) dataset is fit by the resulting (time-dependent) model at the time moment $t_1 = 3.6 \ R_{\rm b}/c$ (time elapsed after the acceleration onset), and the dataset during the peak with the same model at $t_2 = 12.8 \ R_{\rm b}/c$, with the time difference $t_2 - t_1$ correctly matching the $\sim 8.9$ h (observer's frame) interval between the two observations. The best-fit (total) duration of the perturbation, $t_{\rm dur,FII} = t_2 = 12.8 \ R_{\rm b}/c$, translates to $\approx 12.3$ h in the observer's frame, and is well compatible with the observed flux rise time. We therefore connected the low state, the state in the initial stage of the flare, and the state during the flare peak, in a single model within a self-consistent time-dependent framework. The modeling results for the two high states are depicted in Fig.~\ref{fig:modelingflare}. Our model offers a generally reasonable description of the observed data. However, while it reproduces the overall $\gamma$-ray emission level, there is a notable underprediction of the flux below 1 GeV. This discrepancy likely arises from our simplified treatment of the BLR photon field, which only considers a single emission line instead of the actual broader spectrum. Additionally, a contribution from the dusty torus field may be necessary, particularly for $\gamma$-ray production at lower energies, as suggested by previous studies \citep[e.g.][]{hayashida2012, dermer2014}.

It is worth noting that, while blazar emission models often feature parameter degeneracies, our specific model exhibits relatively independent parameters. In particular, the Doppler factor regulates both synchrotron and $\gamma$-ray peak fluxes, with a different scaling relation (which may be somewhat complex due to strong cooling effects), while the magnetic field directly controls the synchrotron peak only. The acceleration time-scale influences the resulting spectral index (``hardness'') in a given spectral band, and the escape time-scale affects the total number of particles in the system, $N_{\rm tot}$. This number has a direct impact on the (relative) level of X-ray flux, which in our scenario is produced mostly due to SSC process, for which the emission is $\propto N_{\rm tot}^2$, rather than $\propto N_{\rm tot}$ as for synchrotron or external Compton processes. This approach yields a seemingly unique solution within our chosen physical scenario.

Comparing with alternative models for this flare, \cite{hess2019flare3c279} model the peak SED within the leptonic external Compton model by a step-change in four parameters, namely the magnetic field, electron injection luminosity, minimum electron Lorentz factor and index of the electron spectrum. Importantly, our scenario, while also incorporating a decrease in the magnetic field, allows the electron spectrum to naturally evolve due to the Fermi-II re-acceleration process, providing a more self-consistent approach.

\begin{figure*}[t]
     \centering
     \begin{subfigure}[b]{0.49\textwidth}
         \centering
         \includegraphics[width=\textwidth]{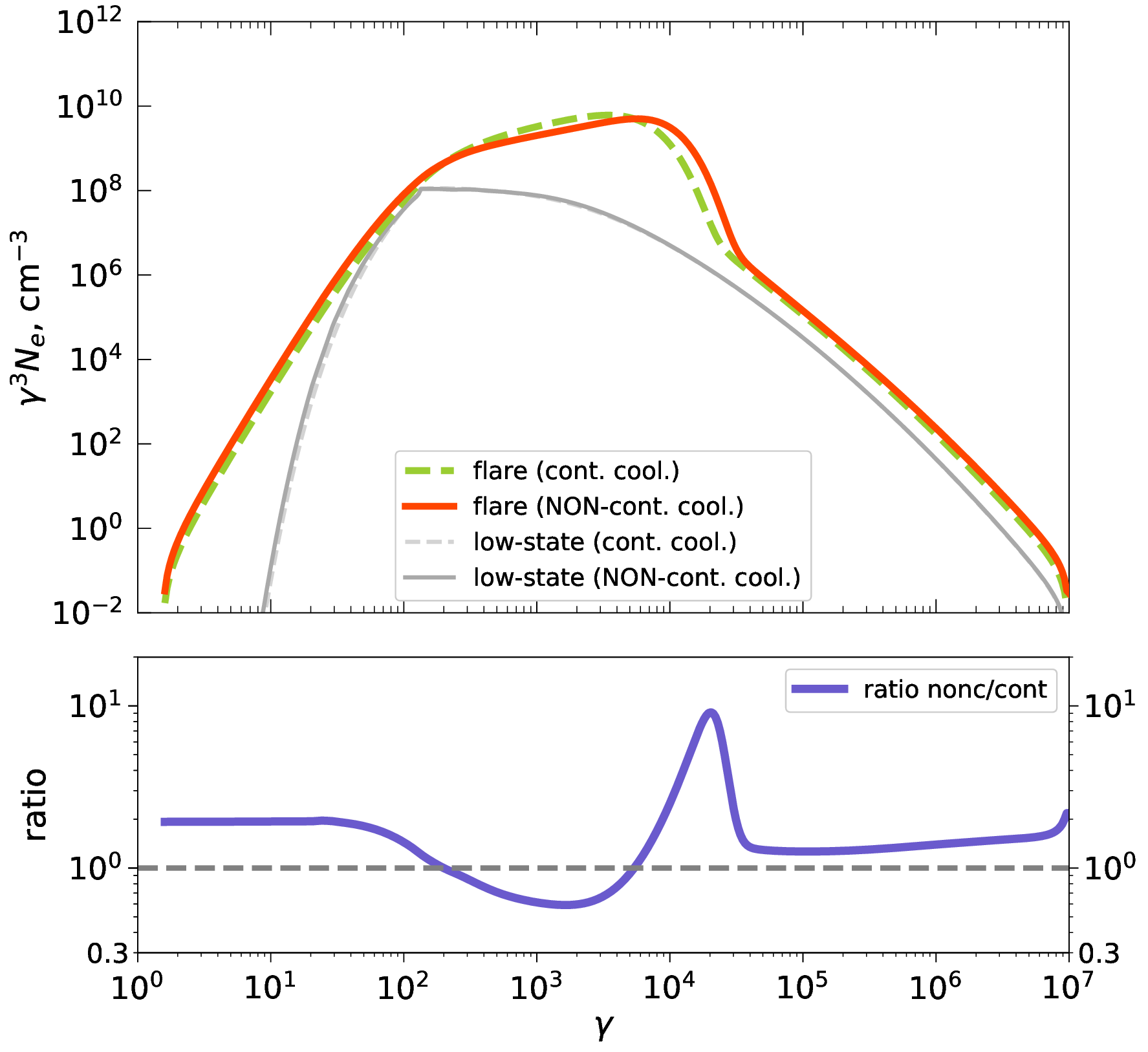}
         \caption{Electron spectrum}
         \label{fig:flarene}
     \end{subfigure}
     \hfill
     \begin{subfigure}[b]{0.49\textwidth}
         \centering
         \includegraphics[width=\textwidth]{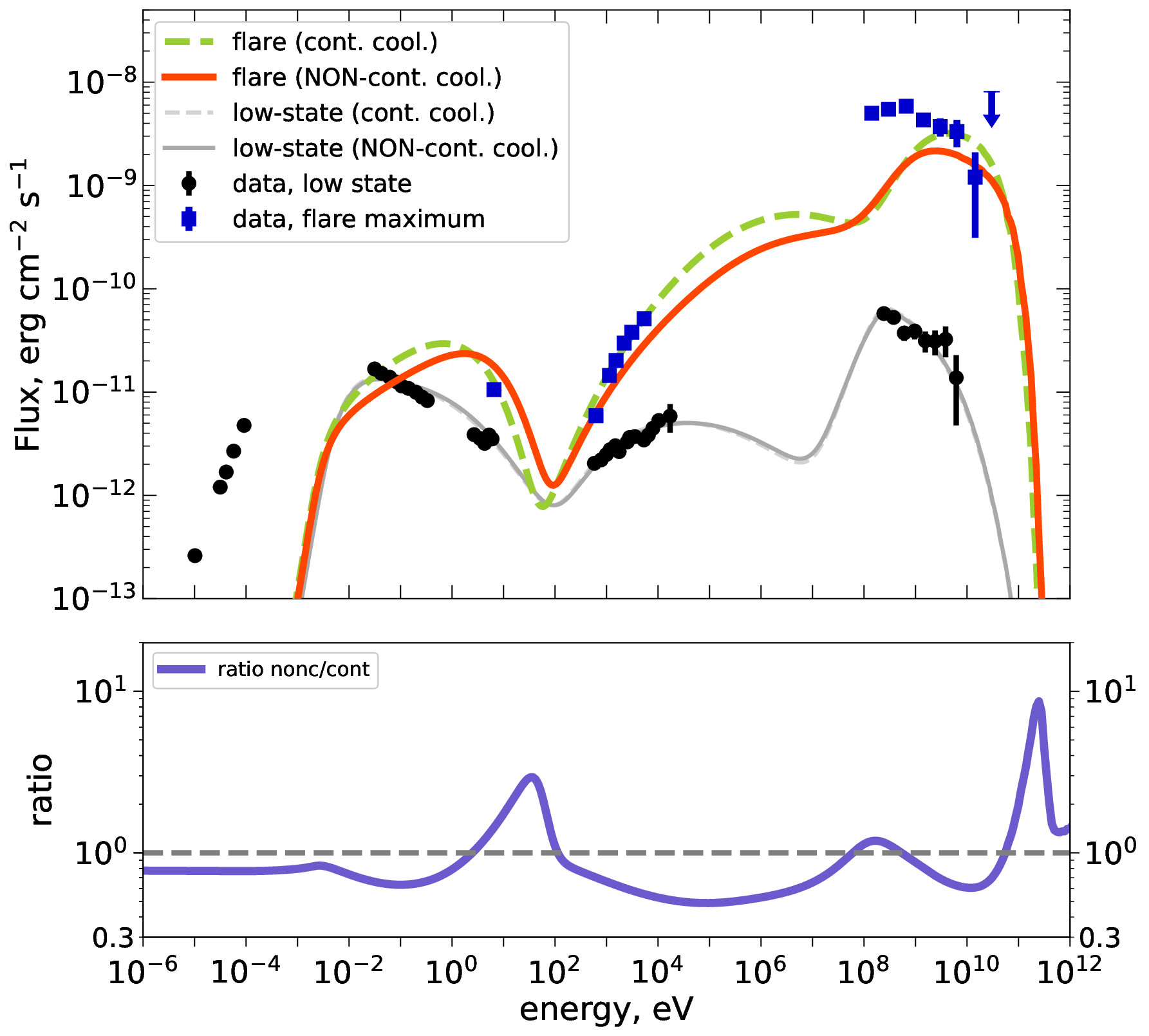}
         \caption{SED}
         \label{fig:flaresed}
     \end{subfigure}
        \caption{Comparison of the electron spectrum (left) and the associated SED (right) between two cooling descriptions for the peak of the June 2015 flaring state of 3C\,279. The electron spectra here are again shown in $\gamma^3 N_e(\gamma)$ representation. The bottom panel of the electron spectrum / SED plot displays the ratio between the electron spectra / SEDs in the case of non-continuous cooling (orange solid curves) and the continuous-loss approximation (green dashed curves) in linear scale.}
        \label{fig:flare}
\end{figure*}

Subsequent to this, we repeat all the above-mentioned simulations of both the low and flaring state using the derived parameters, this time incorporating the full (non-continuous) IC cooling description. A comparison of the underlying electron spectra and SEDs between the non-continuous and continuous-loss scenarios for cooling for the low state is presented in Fig.~\ref{fig:low}, and for the flaring state in Fig.~\ref{fig:flare}.

\section{Discussion}

Comparing the two cooling scenarios in the low state allows us to reveal steady-state implications of non-continuous effects, in a configuration that results from a balance of physical processes. In contrast, the flaring state comparison uncovers dynamic effects arising from transient changes in source conditions and interplay between various physical processes.

\subsection{Low state}

The comparison for the low state (Fig.~\ref{fig:low}) reveals general agreement in electron spectra for Lorentz factors $\gamma \gtrsim 100$. Discrepancies are below $\sim$10\%, with a moderate pile-up observed around the Lorentz factor corresponding to the KN transition, $\gamma \approx 2000$ (Fig.~\ref{fig:lowne}). In the non-continuous cooling scenario (solid curves), the electron spectrum appears harder in the injection-driven regime, showing a ``turning point'' at $\gamma \approx 600$, aligning closely with the anticipated position of the cooling break. Substantial effects emerge below $\gamma_{\mathrm{inj,c}} = 130$, where no injection takes place, including a prominent ``tail'' of particles below $\gamma \sim 10$ and a particle excess above the tail ($10 \lesssim \gamma \lesssim 130$).

Collectively, the discrepancies in the electron spectra result in differences in the observed SED of up to $\lesssim 10$\% (top panels of Fig.~\ref{fig:lowsed}). The most prominent features in the SED for the non-continuous cooling case (compared to continuous-loss case) include a hardening of the spectral slope in the $\gamma$-ray band beyond the peak of the high-energy bump.

The hardening of the electron spectrum around the cooling break, $\gamma_{\rm cool} \simeq (3 m_{\rm e} c^2)/(2 \sigma_{\rm T} R_{\rm b} [u^{\prime}_{\rm rad} + u^{\prime}_{\rm B}])$, together with the excess immediately above around the KN transition, $\gamma_{\rm KN} x \sim 1$, can be attributed to the fact that the continuous-loss approximation might provide an inadequate description of particle cooling in the KN transition regime, tending to overestimate the cooling effect. This inaccuracy, in turn, might arise from the discrete nature of particle cooling. At the same time, this approximation overlooks large relative jumps from the region around $\gamma_{\rm KN}$ to significantly lower Lorentz factors, that a fraction of particles experiences while interacting in the KN regime. This results in an excess and a ``tail'' feature at very low energies. We conclude therefore that the ``tail'' feature might represent an ``echo'' of the particle cooling close to KN transition regime.

\subsection{Flaring state}

An immediate observation is the substantial increase in discrepancies between the two cooling scenarios in the flaring state compared to the low state. One can notice that the low-energy tail and narrow excess observed in the low-state electron spectrum are now replaced by an extended ``shelf''-like excess by a (roughly) constant factor of $\sim 2$, stretching over the Lorentz factors below $\gamma \sim 100$ (Fig.~\ref{fig:flarene}). Below the expected Lorentz factor of the KN transition ($\gamma_{\rm KN} \approx 1300$), a spectral softening is evident, transitioning to spectral hardening above this point. Subsequently, an extreme and relatively narrow pile-up, reaching a factor of $\sim 10$, emerges around $\gamma \sim 2 \times 10^4$. Finally, at the highest Lorentz factors, we observe deviations by a factor $\sim$ 1.3 -- 2.   

These deviations induce strong differences in the SED across the entire frequency domain, with broad-range deviations of up to $\simeq$50\%, along with two narrow pile-ups by a factor of $\sim 3$ and $\sim 8$ (Fig.~\ref{fig:flaresed}). The most substantial discrepancies are observed in the UV, X-ray and VHE bands, along with notable disparities also in the GeV $\gamma$-ray band. Similarly to the low state, the SED in the non-continuous cooling case exhibits a harder slope past the peak of the high-energy bump (GeV-to-VHE regime). Remarkably, the non-continuous cooling model predicts, on average, a several-fold higher flux in the VHE regime during the flare. Upon examination of the SEDs, it is evident that the model SED for the non-continuous cooling case can no longer adequately describe the data. Specifically, the model SED for the non-continuous cooling case exhibits a markedly softer spectral slope in the X-rays, with discrepancies surpassing the (relatively small) uncertainties in the available X-ray data by a very large margin. Important deviations are also observed in the optical-UV band. In the GeV $\gamma$-ray band, discrepancies exceed the {\it Fermi}-LAT uncertainties, particularly in the lower energy range of approximately 1 -- 5 GeV. The non-continuous cooling effects therefore appear to be significant.

Now we proceed to interpretation of the observed features and discrepancies. The substantially enhanced discrepancies between the two cooling scenarios primarily arise from an increase in the overall number of particles in the flaring state (as compared to the low state), as well as an increase in relative number of particles at higher Lorentz factors, particularly beyond the Klein-Nishina threshold ($\gamma > \gamma_{\rm KN}$), as a result of the re-acceleration process. Additionally, a slightly increased Lorentz and Doppler factor during the flaring state leads to a boost of the seed photon field in the blob frame and amplification of the cooling rate, as well a boost in the target photon energy. Consequently, the majority of the $\gamma$-ray production now occurs much closer to the KN regime, enhancing the discrete cooling effects. 

For the domain below $\gamma_{\mathrm{inj,c}}$, unlike in the low state, it is now no longer dominated by losses only (cooling and escape). Instead, Fermi-II acceleration now competes with cooling effects. With the inferred Fermi-II time-scale, $3.6 \ R_{\rm b}/c$, acceleration dominates for $\gamma \leq 50$, i.e.\ $t_{\rm cool, syn+IC}(\gamma) > t_{\rm acc,FII}$, with $t_{\rm cool, syn+IC}(\gamma) = \gamma/(\dot{\gamma}_{\rm IC} + \dot{\gamma}_{\rm syn})$. Hence, the ``shelf''-like feature below $\gamma \sim 100$ can be attributed to the Fermi-II process re-accelerating low-energy particles downscattered from interactions in KN regime, effectively spreading them into an extended homogeneous excess.

Above the KN transition, as the IC cooling loses its efficiency, the total cooling time-scale increases with a higher $\gamma$ and eventually becomes comparable to the acceleration time-scale again ($t_{\rm cool, syn+IC} \sim t_{\rm acc,FII}$), occurring around $\gamma \approx 3 \times 10^4$, where the cooling time-scale exhibits a local maximum. Beyond this point, synchrotron cooling begins to dominate the total cooling rate, with the cooling time-scale decreasing again with an increasing $\gamma$, resulting in synchrotron cooling strongly prevailing above the acceleration process ($\gamma > 10^5$). In the Lorentz factor range where cooling is the most inefficient ($\gamma \sim 10^4$ -- $10^5$), a distinct ``bump'' appears in the electron spectra, due to maximized relative importance of re-acceleration. The position of this bump for the non-continuous cooling scenario is appreciably shifted towards higher Lorentz factors. This can be explained by the continuous-loss approximation potentially underestimating the decline in the IC process efficiency in the KN regime, i.e.\ overestimating the overall cooling effect. The shift in the position of the bump leads to the observed strong pile-up at $\gamma \approx 3 \times 10^4$, as well as the pronounced narrow pile-up appearing in the low- and high-energy SED component. At the highest Lorentz factors, the electron spectrum for the non-continuous cooling case consistently appears higher (showing an extended particle excess), likely due to the same reason of overestimation of the cooling effect by the continuous-loss scenario.

\section{Conclusions}

In this work, we have for the first time explored in detail the non-continuous (discrete) IC cooling effects in $\gamma$-ray (Compton) dominated blazars (FSRQs), arising due to IC interactions proceeding in the KN regime. Our analysis centers on the archetypal FSRQ 3C\,279, and involves modeling of the brightest $\gamma$-ray flare of this source observed in June 2015. We find that the discrete cooling effects may become important and significantly affect the electron spectra and SEDs of blazars, inducing a range of discerning features, such as low-energy tails, narrow and extended excesses / pile-ups, spectral softening and hardening, shifts in the position of cooling bump/break, etc. In more detail, our study has unveiled the following main effects:

   \begin{enumerate}

    \item {\bf Low State}: In the injection-dominated regime, the electron spectra agree well with discrepancies below 10\%. The most notable features emerge at low Lorentz factors in the injection-free regime, and include a significant particle excess and a low-energy ``tail'', which appear to be low-energy reflections of the KN transition. We interpret these features as due to particles interacting in the KN regime and experiencing large jumps to much lower Lorentz factors, an effect that is not properly taken into account by the continuous-loss description. These deviations lead to minor SED differences of up to 10\%. Therefore, non-continuous cooling effects can be neglected in the low-states of blazars.

    \item {\bf Flaring state}: During flares, the electron spectrum and SED show significantly amplified discrepancies between continuous and non-continuous cooling, yielding a distinct array of features. Specifically, we investigated a (typical) flare scenario, where the flux increase is initiated by transient particle re-acceleration (turbulent or Fermi-II mechanism in this case), with acceleration competing with cooling. In the flaring state, a pronounced extended particle excess emerges at low Lorentz factors, stemming from particles downscattered to very low Lorentz factors and subsequently re-accelerated. Furthermore, a moderate excess at high Lorentz factors, as well as a significant shift in the position of cooling ``bump''/break (resulting from a competition between acceleration and different regimes of cooling), suggest an overestimation of cooling effects in the KN regime by the continuous-loss approximation. Broad-range SED deviations peak at 50\%, with the most prominent ones seen in the hard X-ray and VHE band. Also, the shape of the high-energy SED bump ($\gamma$-ray domain) displays a noticeable distortion, including a considerable hardening of the SED in the GeV-to-VHE domain in the non-continuous case, as well as a strong narrow pile-up. The discrepancies between the two cooling scenarios are found to exceed the uncertainties of the observational data by a considerable margin, notably in the X-ray domain. This highlights the importance of the non-continuous effects during flares and underscores that their imprint on the emission spectra can be effectively discerned in the modeling of data from current MWL instruments, while even more so with future-generation observatories like CTA.

    \end{enumerate}

In summary, taking into account the non-continuous cooling effects is essential for accurate modeling of blazar flaring states, as these effects become non-negligible during strong FSRQ flares characterized by high Compton dominance.

\begin{acknowledgements}
The work of M.B. was supported by the South African Research Chairs Initiative of the National Research Foundation\footnote{Any opinion, finding, and conclusion or recommendation expressed in this material is that of the authors, and the NRF does not accept any liability in this regard.} and the Department of Science and Innovation of South Africa through SARChI grant no. 64789. A.D. thanks Andrzej Zdziarski for useful discussions on the subject. A.D. acknowledges the FSK-GAMMA-1 computer cluster facility at the Centre for Space Research (North-West University) which was particularly helpful for performing computationally-expensive simulations for this project. Also, A.D. thanks Pieter Van der Merwe and Robert Brose for their ideas and suggestions that were of great help for the numerical implementation of the full transport equation and for the code optimization. Finally, the authors extend their gratitude to the anonymous referee for their critical and thorough review of the manuscript, as well as for providing important requests and suggestions that significantly enhanced the quality of the presented work.    
\end{acknowledgements}

%
%

\bibliographystyle{aa} 
\bibliography{cooling_biblio}

\end{document}